\documentclass[american,english]{article}
\usepackage[T1]{fontenc}
\usepackage[latin9]{inputenc}
\usepackage{geometry}
\usepackage{color}
\usepackage{babel}
\usepackage{refstyle}
\usepackage{booktabs}
\usepackage{units}
\usepackage{amsmath}
\usepackage{graphicx}
\usepackage[numbers]{natbib}
\usepackage[unicode=true,pdfusetitle,
 bookmarks=true,bookmarksnumbered=false,bookmarksopen=false,
 breaklinks=false,pdfborder={0 0 1},backref=false,colorlinks=true]
 {hyperref}
\hypersetup{
 linkcolor=magenta, urlcolor=blue, citecolor=blue, pdfstartview={FitH}, hyperfootnotes=false, unicode=true}
\usepackage{breakurl}

\makeatletter

\AtBeginDocument{\providecommand\figref[1]{\ref{fig:#1}}}
\AtBeginDocument{\providecommand\subsecref[1]{\ref{subsec:#1}}}
\AtBeginDocument{\providecommand\eqref[1]{\ref{eq:#1}}}
\AtBeginDocument{\providecommand\tabref[1]{\ref{tab:#1}}}
\providecommand{\tabularnewline}{\\}
\RS@ifundefined{subsecref}
  {\newref{subsec}{name = \RSsectxt}}
  {}
\RS@ifundefined{thmref}
  {\def\RSthmtxt{theorem~}\newref{thm}{name = \RSthmtxt}}
  {}
\RS@ifundefined{lemref}
  {\def\RSlemtxt{lemma~}\newref{lem}{name = \RSlemtxt}}
  {}

\newcommand{\lyxaddress}[1]{
\par {\raggedright #1
\vspace{1.4em}
\noindent\par}
}

\usepackage{algorithm,algpseudocode}

\makeatletter
\newcommand*{\centerfloat}{%
  \parindent \z@
  \leftskip \z@ \@plus 1fil \@minus \textwidth
  \rightskip\leftskip
  \parfillskip \z@skip}
\makeatother

\usepackage{caption}
\usepackage{booktabs, multirow}
\usepackage{hyperref}

\makeatother

\begin{document}

\title{Efficient low-dimensional approximation of continuous attractor networks}

\author{Alexander Seeholzer$^{\text{1}}$, Moritz Deger$^{\text{1,2}}$,
Wulfram Gerstner$^{\text{1}}$}
\maketitle
\selectlanguage{american}%

\lyxaddress{\noindent {\small{}1: School of Computer and Communication Sciences
and School of Life Sciences, Brain Mind Institute, }\\
{\small{}\'Ecole Polytechnique F\'ed\'erale de Lausanne, 1015 Lausanne
EPFL, Switzerland}\\
{\small{}2: Institute for Zoology, Faculty of Mathematics and Natural
Sciences, }\\
{\small{}University of Cologne, 50674 Cologne, Germany}}
\selectlanguage{english}%
\begin{abstract}
Continuous ``bump'' attractors are an established model of cortical
working memory for continuous variables and can be implemented using
various neuron and network models. Here, we develop a generalizable
approach for the approximation of bump states of continuous attractor
networks implemented in networks of both rate-based and spiking neurons.
The method relies on a low-dimensional parametrization of the spatial
shape of firing rates, allowing to apply efficient numerical optimization
methods. Using our theory, we can establish a mapping between network
structure and attractor properties that allows the prediction of the
effects of network parameters on the steady state firing rate profile
and the existence of bumps, and vice-versa, to fine-tune a network
to produce bumps of a given shape.
\end{abstract}
\emph{A previous version of this article was published as Chapter
3 of the first author's Ph.D. thesis \citep{Seeholzer2017}}

\section{Introduction}

Behaving animals commonly need to transiently memorize information
about the environment. For example, as an animal overlooks the visual
scenery, locations of certain salient stimuli need to be recorded
and stored. Such information does not need to be stored in long-term
memories. Rather, working memory must provide a quickly accessible
computational substrate for storing information over short durations.
While long-term memory is thought to be stored in the efficacy of
synaptic connections in the brain \citep{Hebb1949,Martin2000,Takeuchi2014},
a possible substrate for working memory may be transiently stable
states of neuronal activity across cortical networks \citep{Goldman-Rakic1995,Curtis2003,Chaudhuri2016}.

As model implementations of this concept, localized spatial profiles
of neural activity have been proposed for the internal representation
of sensory stimuli \citep{Wilson1973,Amari1977,Ben-Yishai1995}. First,
neurons are associated to the presence of physical quantities through
elevated responses during and after the presentation of stimuli, akin
to receptive fields. For example, the presentation of stimuli at varying
angular positions in the visual field evokes persistent and elevated
firing rates in selective groups of neurons of the prefrontal cortex
during delay periods \citep{Funahashi1989}. Choosing recurrent connection
weights (or connection probabilities) which are stronger between neurons
that are responsive to similar stimuli, together with feedback inhibition
limiting the total firing rates in a network, allows this class of
models to display \emph{bumps} of self-sustained activity: neuronal
activity that is localized in the space of possible stimuli. Since
these states are stable attractive states, and all possible such states
form a continuum, these models are often referred to as \emph{continuous
attractors}. The elevated firing of neurons responsive to similar
stimuli is then seen as the working memory representation of physical
quanta stored in the network, e.g. spatial orientations \citep{Zhang1996},
or angular positions in the visual field \citep{Compte2000}. Similar
computational circuits might also serve as the basis of persistent
internal representations in hippocampal areas \citep{Itskov2011,Yoon2013a}. 

Continuous attractor models with simplified shapes of connectivity
or neuronal input-output relations can be analyzed and often exactly
solved \citep{Wilson1973,Amari1977,Ben-Yishai1995,Bressloff2003,Fung2010,Itskov2011a,Laing2014},
or may generally be approximated in the linear input-output regime
of balanced networks \citep{Rosenbaum2014}. However, the inclusion
of biologically plausible nonlinearities, like nonlinear neuronal
input-output relations \citep{Compte2000,Renart2003}, neuronal adaptation
\citep{Brette2005,Roach2015}, or synaptic nonlinearities like short-term
plasticity \citep{Itskov2011a,Zucker2002} and saturating NMDA kinetics
\citep{Compte2000,Destexhe1994}, complicate the mathematical solution
of these systems considerably and make a derivation of the stable
firing rate profile unfeasible. Therefore, such systems are usually
studied by explicit simulations of the underlying dynamics or by numerical
optimization of approximated equations for all neurons \citep{Rosenbaum2014,Spiridon2001}.
While these procedures in principle allow the prediction of the network
activity as a function of the parameters, they involve computationally
demanding numerical optimization of high-dimensional systems of equations,
possibly as costly as simulating the full neuronal dynamics. Thus,
currently, relating the microscopic network parameters to the resulting
emergent bump states involves repeated and possibly time consuming
simulations. For example, this makes the matching of the network steady
states to physiologically constrained features tedious.

Here, we present a generalizable approach for the approximation of
the network-wide steady states of continuous attractors. Our approach
allows the prediction of the shape of steady-state firing rate profiles,
under nonlinear neuronal dynamics and varying configurations of the
underlying microscopic system, without having to solve the dynamics
of the full, high-dimensional system. Our novel method relies on a
low-dimensional parametrization of the network's firing rate profile,
which allows us to derive computationally tractable systems of self-consistent
equations describing the attractor steady-state, akin to mean-field
approaches for networks with discrete attractors \citep{Brunel2001}.
These equations can be used to efficiently predict the dependence
of the firing rate profile on microscopic network parameters. Importantly,
because the dimension of the parameterization of the spatial activity
profile is low, our approach makes optimization of the microscopic
network parameters for the appearance of desired bump profiles feasible.
We apply our method to both networks of simplified rate neurons, and
networks of complex, conductance-based integrate-and-fire neurons
with saturating and voltage-dependent nonlinear NMDA transmission.

\section{Results}

Mean-field approaches (see e.g. \citep{Brunel2001,Amit1997}) that
predict the steady states of recurrently connected neuronal networks
usually rely on dimensionality reduction. The number of equations
describing the dynamics is reduced by partitioning neurons into groups
of ``similar'' neurons, and deriving expressions which describe
the average statistics for these coupled groups in the steady states.
For example, the simplest such partition consists in considering the
mean firing rates of excitatory and inhibitory neurons separately,
e.g. all excitatory neurons fire with similar mean rates given the
same input:
\[
\nu_{E}=F_{E}\left(\text{input}_{E}\right).
\]

If the groups of neurons are now homogeneously coupled, i.e. the connections
between neurons depend only on the groups of the neurons involved,
one can derive the input to neurons of each group in dependence of
the firing rates of the groups only. This leads to a closed system
of self-consistency equations describing the coupled steady-state
firing rates:
\begin{align*}
\nu_{E} & =F_{E}\left(\text{input}_{E}\left[\nu_{E},\nu_{I}\right]\right),\\
\nu_{I} & =F_{I}\left(\text{input}_{I}\left[\nu_{E},\nu_{I}\right]\right).
\end{align*}

In the steady states of continuous attractor models (see \figref{1}),
neurons fire at different rates, making a clear partition into discrete
groups of similarly firing neurons difficult. Therefore, the solution
of such systems usually relies on the explicit simulation of the neuronal
dynamics of all neurons along the spatial dimension, or a numerical
solution of the coupled self-consistency equations for all neurons.

Here, by using the continuity of the shape of the attractor states,
we demonstrate that continuous attractors are amenable to dimensional
reduction, by parametrizing the attractor state by a low-dimensional
family of functions. In \subsecref{Approximation-of-ring-attractor},
we check our method on networks of simple rate neurons, for which
the method might not yield much improvement over simulations or numerical
solutions of the steady states. For the spiking networks considered
in \subsecref{Spiking-networks}, we show that our approach speeds
up predictions of the steady states considerably and further makes
these networks amenable to the optimization of network parameters.

\subsection{\label{subsec:General-mean-field-equations}General equations for
the approximation of stable states in ring-attractors}

\begin{figure}
\includegraphics[width=1\linewidth]{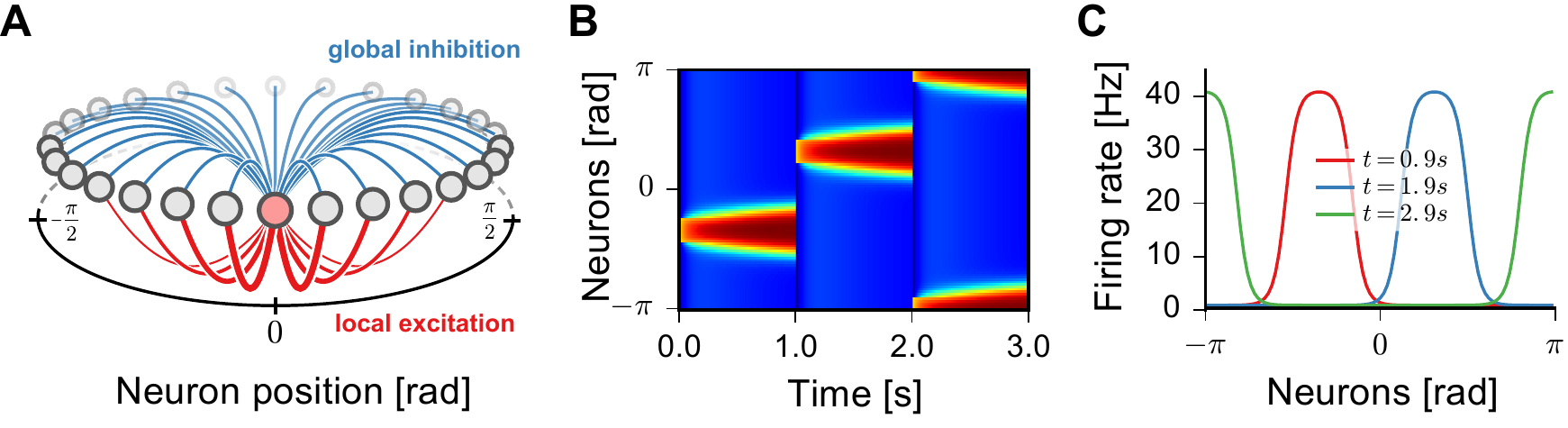}\caption[Stable firing rate profiles in continuous attractors]{\textbf{\label{fig:1}Stable firing rate profiles in continuous attractors.}
\textbf{A:} Neurons (circles) are assigned a position on a ring and
connected with distance-dependent connection weights to all other
neurons: the firing of a neuron (red circle) inhibits all other neurons
(global inhibition, blue lines) and strongly excites neurons close
it (local excitation, red lines). \textbf{B:} Example simulations
of a continuous attractor. At times $t=0,1,2s$ the system is reset
to localized activity centered at different positions. With time the
activity bump broadens towards a stable state. Colors indicate firing
rates as in panel C.\textbf{ C:} Firing rates of the network shown
in panel B, measured close to the stable states at $t=0.9,1.9,2.9s$.
Plots are generated using the rate model introduced in \subsecref{Rate-network}.
}
\end{figure}
As a concrete class of continuous attractors, we consider the ring-attractor
model, in which stable bumps of neuronal activity are freely translatable
along all positions on a circle. Ring-attractor models can be constructed
by placing $N$ neurons (rate-based or spiking) at equally spaced
angular positions $\theta$ along the ring (\figref{1}A) \citep{Wilson1973,Amari1977,Ben-Yishai1995}.
We choose the angular space to consist of positions $\theta\in[-\pi,\pi)$,
where we identify the ends of the interval: a neuron at position $\theta=\pi-\epsilon$
is the neighbor of a neuron at position $\theta=-\pi.$ At short angular
distances, recurrent connections are chosen to be strong and excitatory,
while neurons further apart in angular space effectively inhibit each
other's firing (\figref{1}A). Due to the symmetry in connectivity
with respect to distance, these networks can form a continuous manifold
of stable states for sufficiently strong connections: the network
activity in response to external inputs converges to firing rate profiles
centered around some angular position. The position can be, for example,
controlled by providing an external input to the network centered
around any desired position (\figref{1}B). The stereotypical shape
of the resulting firing rate profiles (\figref{1}C) is invariant
with respect to translations in angular space.

\begin{figure}
\includegraphics[width=1\linewidth]{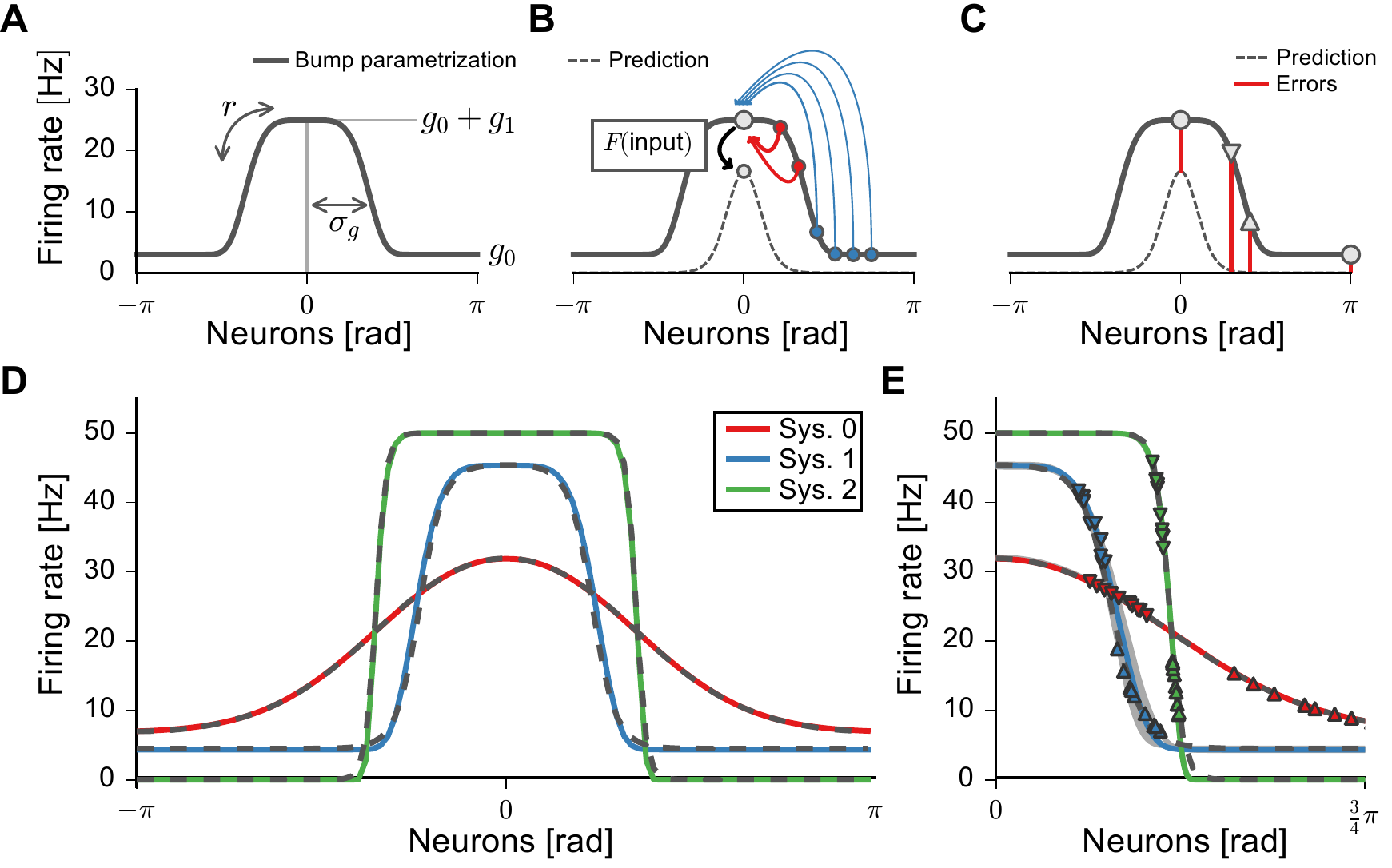}\caption[Approximation of bump shapes in attractor networks]{\textbf{\label{fig:2}Approximation of bump shapes in attractor networks.}
\textbf{A:} Parametrization of firing rate profiles by 4 shape parameters:
$g_{0}$ and $g_{1}$ control the baseline and maximal firing rates,
$\sigma_{r}$ controls the width, and $r$ controls the steepness.\textbf{
B: }Assuming a spatial profile given by the parametrization $g$,
the input to any neuron can be calculated by summing all synaptic
inputs: in the example, the neuron at position $0$ receives some
strong local excitatory input (red thick arrows) and weaker inhibitory
input from more distant neurons (blue thin arrows). The firing rate
prediction (dashed line) for any neuron can then be calculated as
a function of its input. This is illustrated for a neuron at $0$
(circle). \textbf{C:} Self-consistency errors between the current
parametrization (gray line) and the resulting firing rate prediction
(dashed line) are calculated at a small number of points along the
firing rate profile. Intermediate points (triangles) are positioned
dynamically during optimization. \textbf{D:} Optimized bump parametrization
(dashed lines) for systems with three different connectivities (solutions
of the full system are plotted in color). \textbf{E:} Dependence of
solution on intermediate point placement. Left points are given by
downward triangles, rightward points by upward triangles (compare
to panel \textbf{C}). Single optimization runs (light gray lines)
together with median parameters (dashed lines, same as in D) and full
solutions (colored lines). }
\end{figure}

The continuity of the firing rate profile allows us to parametrize
the firing rates in the population by a small number of parameters,
and to derive equations from the underlying model that constrain these
parameters. Here, inspired by shapes observed in simulations, we choose
to parameterize the firing rate profile by a generalized Gaussian
function, where we assume without loss of generality that the distribution
is centered at $\theta=0$ (cf. \figref{2}A):
\begin{equation}
g(\theta)=g_{0}+g_{1}\exp\left(-\left[\frac{\left|\theta\right|}{g_{\sigma}}\right]^{g_{r}}\right).\label{eq:g-gen-gaussian}
\end{equation}
Here, $g_{0}$ controls the baseline firing rate and $g_{0}+g_{1}$
will be the maximum firing rate of the profile. The parameters $g_{\sigma}$
and $g_{r}$ control the width and steepness of the profile, respectively.
If we know the distance-dependent connectivity $w$ between neurons
and their input-output relation $F$, we can predict the expected
neuronal firing at any position $\theta$ in the population (cf. \figref{2}B).
Crucially, we want the firing rate profile $g(\theta)$ to be generated
by the neuronal dynamics \textendash{} we thus identify $g(\theta)$
at the point $\theta$ with the firing rate $\nu(\theta)$ of a neuron
at position $\theta$. Finally, we replace the synaptic input to the
neuron at position $\theta$ with the contributions from all neurons
firing at rates $g$ along the ring. For any given position $\theta_{i}$
along the ring, this yields a self-consistent equation in the function
$g$:
\begin{align}
g(\theta_{i}) & =F(\text{input}_{\theta_{i}}\left[g\right])\nonumber \\
 & =F\left(\int_{-\pi}^{\pi}d\varphi w(\varphi-\theta_{i})g(\varphi)\right),\label{eq:mfeq-selfconsist}
\end{align}
with the corresponding \emph{self-consistency error}
\begin{align}
\text{Err}_{i} & \equiv g(\theta_{i})-F(\text{input}_{\theta_{i}}\left[g\right]).\label{eq:error-func}
\end{align}

In principle, this procedure can yield up to $N$ coupled error functions,
one for each of the $N$ neurons. One could then minimize the quadratic
error $\sum_{i}\text{Err}_{i}^{2}$ with respect to the parameters
$\{g_{0},g_{1},g_{\sigma},g_{r}\}$ to find an approximate solution
of the system. However, since the evaluation of each error function
can be costly (e.g. in spiking networks, see \subsecref{Prediction-of-firing}),
we propose a low-dimensional approximation to constrain the set of
$4$ free parameters: we pick only $4$ points $\theta_{1},\dots,\theta_{4}$,
at which we evaluate the errors. This assumes that the shapes of firing
rate profiles maintained by the network are well approximated by the
function $g(\theta)$, which we found to be the case for all networks
considered here. This leaves the choice of points $\theta_{i}$ to
evaluate. To ensure that errors are evaluated across different firing
rates, we set the position of these points to cover a range of function
values $h_{i}=g(\theta_{i})$: we choose the top of the distribution
$\theta_{1}=0$ with $h_{1}=g_{1}+g_{0}$, as well as the lowest point
$\theta_{4}=\pi$ with $h_{4}=g_{0}$ (circles in \figref{2}C). The
remaining intermediate points (triangles in \figref{2}C) are dynamically
positioned (see \subsecref{Placement-of-sampling} in Methods): their
position depends on the function $g$ such that they always sample
intermediate function values $h_{i}$.

\subsection{\label{subsec:Approximation-of-ring-attractor}Approximation of ring-attractor
profiles in rate models}

The proposed method is, in principle, applicable to any neuron model
with a defined input-output relation $F$. The shape of the stable
attractor profiles will, however, depend on the concrete choice of
neuron model and the microscopic parameters, in particular the parameters
governing the connectivity between neurons. To test the ability of
the low-dimensional approximation proposed in the last section to
correctly predict the shapes of firing rate profiles, we implemented
the ring-attractor model introduced above (see \figref{1}) in a network
of rate-based neurons with $\tanh$ input-output function and a generalized
Gaussian recurrent connectivity (see \subsecref{Rate-network} in
Methods). For the rate neuron models chosen here, the self-consistency
errors \eqref{error-func} are given by (see \subsecref{Derivation-of-input-output-rate}
in Methods): 
\begin{align}
\text{Err}_{i} & =g(\theta_{i})-\frac{\nu_{\text{max}}}{2}\left[1+\tanh\left(\frac{\tau_{s}}{s_{0}}\frac{N}{2\pi}\text{input}_{\theta_{i}}\left[g\right]\right)\right],\nonumber \\
 & =g(\theta_{i})-\frac{\nu_{\text{max}}}{2}\left[1+\tanh\left(\frac{\tau_{s}}{s_{0}}\frac{N}{2\pi}\int_{-\pi}^{\pi}d\varphi w(\varphi-\theta_{i})g(\varphi)\right)\right],\label{eq:rate-error}
\end{align}
where $\tau_{s}$ is the time-constant of synaptic inputs, $\nu_{\text{max}}$
is the maximal firing rate and $s_{0}$ is an input scale.

\subsubsection{Prediction of stable firing rate profiles}

To approximate the firing rate profiles that these networks admit
as self-consistent solutions, we minimize the error functions \eqref{rate-error}
with respect to the parameters $\{g_{0},g_{1},g_{\sigma},g_{r}\}$
as free variables. We find that, for a range of connectivity parameters
(see \tabref{weights-systems}), the predicted shapes converge to
unique solutions. This solution matches the steady state of the microscopic
network simulations accurately (\figref{2}D). This is the case both
for attractor states that lie in the linear regime of the neuronal
input-output relations (\figref{2}D, red line) as well as for highly
nonlinear attractor dynamics in which neuronal firing reaches saturation
values, leading to plateau-shaped firing rate profiles (\figref{2}D,
blue and green lines).

As discussed above, the placement of intermediate sampling points
(\figref{2}C, triangles) is not constrained by theory, but remains
a free parameter of our approach. We chose these points at positions
$\theta_{i}$ such that they sample given function values $h_{i}=g(\theta_{i})$.
To investigate the dependence of the prediction on the placement of
intermediate sampling points, we calculated several predictions while
randomly varying the choices of $h_{i}$ (\figref{2}E, triangles).
We find that this hardly affects the converged solutions.

\subsubsection{Optimization of network parameters}

\begin{figure}
\includegraphics[width=1\linewidth]{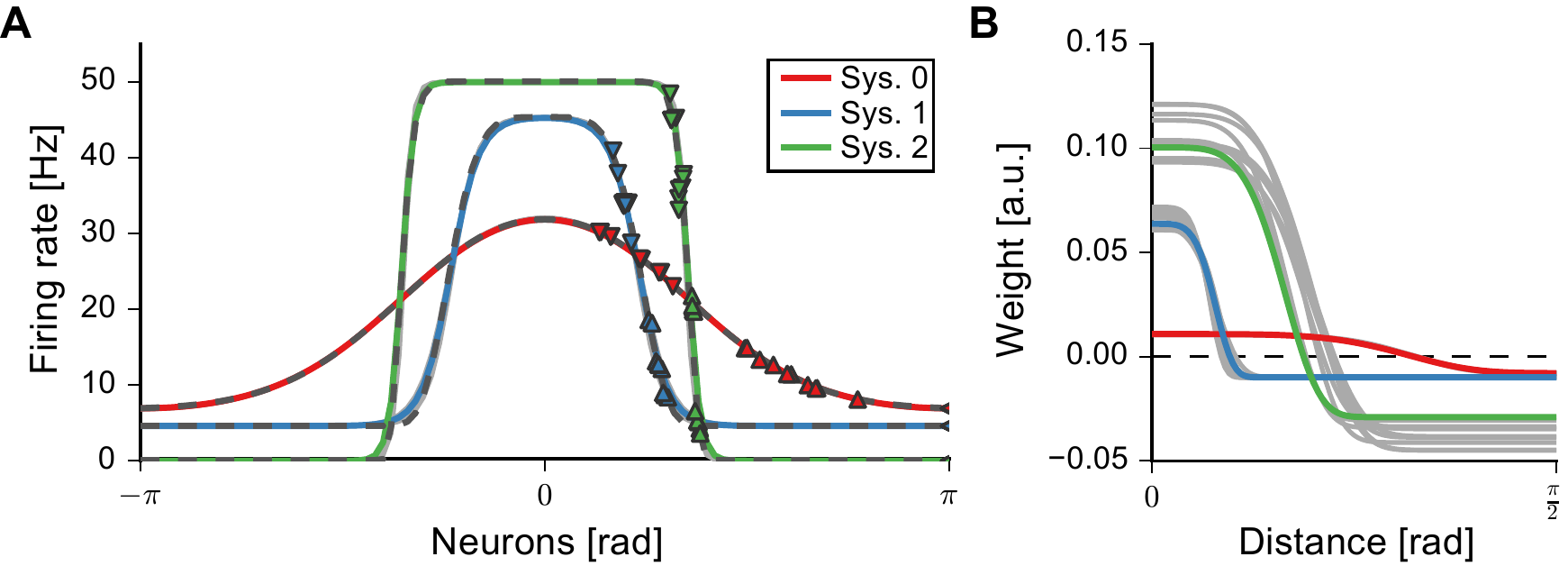}\caption[Application: Optimization of network connectivity]{\textbf{\label{fig:3}Application: Optimization of network connectivity.
A:} Stable network firing rate profiles resulting from 10 optimization
runs of the network local connectivity profiles. The optimization
target profiles are plotted in colors, together with single optimization
runs (light gray lines, almost overlapping with colored lines) and
one example highlighted optimization result (dashed lines) each. Left
optimization points are given by downward triangles, rightward points
by upward triangles (compare to \figref{2}C,E). \textbf{B:} Results
of the 10 optimized connectivity profiles (light gray lines) together
with one example profile (colored lines). Solutions for System $2$
are degenerate, while others are fairly unique. }
\end{figure}
While our low-dimensional system of self-consistent equations can
be used for the prediction of the steady-state firing rates, they
can also be used in the inverse way, to optimize any of the network
parameters. We demonstrate this here, using the shape of the recurrent
connectivity as an example. However, such optimizations can include
further parameters of the network model (see \subsecref{Optimization-of-network}).

To optimize the network connectivity parameters, we keep the parametrization
parameters $\left\{ g_{0},g_{1},g_{\sigma},g_{r}\right\} $ fixed
to the desired values of the shape of the firing rate profile. We
then optimize the self-consistent equations \eqref{mfeq-selfconsist}
for values of the recurrent connectivity parameters $\left\{ w_{0},w_{1},w_{\sigma},w_{r}\right\} $
which lead to solutions of the equations and produce the desired bump
profile. In \figref{3} we show the results of this procedure for
the three systems also investigated in \figref{2}. The procedure
yields network connectivities that fulfill the desired properties
(\figref{3}A), largely independently of the points $\theta_{i}$
chosen for the evaluation of the errors. Importantly, for some shapes
the solutions show a degeneracy (\figref{3}B, gray lines), in the
sense that several connectivity parameter sets are found that produce
the same stable firing rate profile.

\subsection{\label{subsec:Spiking-networks}Approximation of ring-attractor profiles
in spiking networks}

\begin{figure}
\includegraphics[width=1\linewidth]{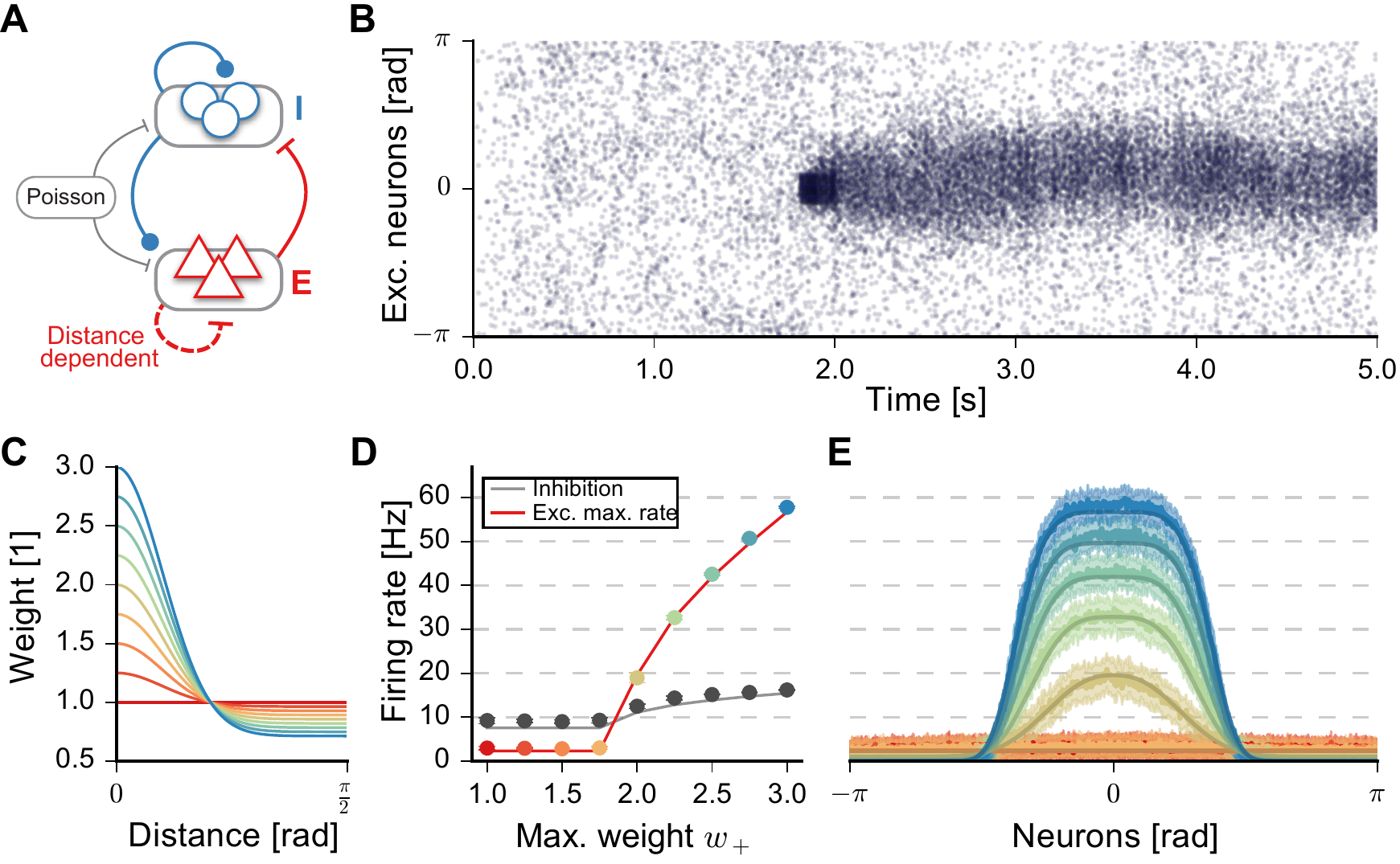}\caption[Application: Prediction of firing rate profiles in spiking continuous
attractor networks]{\textbf{\label{fig:4}Application: Prediction of firing rate profiles
in spiking continuous attractor networks}. \textbf{A:} Recurrently
connected spiking network of a population of excitatory (E, red triangles)
and inhibitory (I, blue circles) neurons. Networks are fully connected
with uniform weights, except for E-E connections (dashed red line),
which are distance-dependent. All neurons receive excitatory input
with spikes generated by homogeneous Poisson processes. \textbf{B:}
Example simulation: E neurons fire asynchronously and irregularly
until an external cue is given (centered at $0$ for $0.2s$ starting
at $t=1.8s$). After this stimulus, a bump of elevated activity sustains
itself around the point of stimulation ($w_{+}=2.0$). \textbf{C:}
Distance-dependent E-E connectivity as a function of the parameter
$w_{+}$ (maximal strength or recurrent connections). Values range
between $w_{+}=1$ (red) and $w_{+}=3$ (blue). \textbf{D:} Appearance
of the bump firing rate profile as a function of the connectivity
parameter $w_{+}$. Theoretical predictions (lines) and simulation
results (maximum of fit of $g$ to firing rates recorded over $1.5$s
of delay activity, mean over $5$ repetitions, errors show $95\%$
CI) for maximal firing rates of E (colored) and I (gray) neurons.
Colors similar to those in panel C.\textbf{ E:} Shape of the optimized
firing rates profiles (solid colored lines) compared to mean firing
rates measured from spiking simulations (thin colored lines) (mean
$\pm0.5\text{STD}$ of firing rates recorded from $1.5s$ of delay
activity in $5$ repetitions) for all values of $w_{+}$ in panel
C and D (similar colors). }
\end{figure}
In complex spiking neuron models, the steady-state input-output relations
often involve integral functions \citep{Brunel2001,Brunel1999} that
are not amenable to further theoretical analysis. Further, the introduction
of nonlinear NMDA transmission at excitatory synapses \citep{Compte2000,Wang1999}
complicates the analysis of such models considerably: voltage-dependent
gating of the maximal NMDA conductance (by the voltage-dependence
of the $\text{Mg}^{2+}$ block) and saturation of the NMDA at high
conductances necessitates the numerical solution of a $4$-dimensional
system of self-consistent equations for a relatively simple 2 population
mean-field model \citep{Brunel2001} (see below).

Here, we demonstrate that firing rate profiles in a continuous attractor
network of spiking neurons with such nonlinear NMDA transmission are
amenable to the same approach as described above, which still involves
evaluating comparatively few equations. The spiking network we implement
(similar to \citep{Compte2000} and used with variations in e.g. \citep{Renart2003,Murray2012,Wei2012,Pereira2015,Almeida2015},
see \subsecref{Spiking-network-methods} in Methods) consists of two
fully connected populations of conductance-based integrate-and-fire
neurons: a population of inhibitory neurons with unstructured all-to-all
connectivity, and a population of excitatory neurons, with distance-dependent
recurrent excitatory connections (\figref{4}A). In addition, all
neurons receive excitatory background input mediated by spikes generated
by Poisson processes. These network can be tuned such that they possess
a bi-stability (\figref{4}B): a uniform state with spontaneous spiking
activity in the excitatory population (the inhibitory population is
always uniformly spiking) coexists with an ``evoked'' spatially
inhomogeneous bump-state that appears after an external cue input
is given to a subgroup of excitatory neurons (stimulus is present
at $t=1.8-2.0s$ in \figref{4}B).

\subsubsection{Self-consistent equations for networks of spiking neurons}

For the excitatory population, we again parametrize the spatial profile
of firing rates by \eqref{g-gen-gaussian}, which allows us to derive
self-consistent equations for any neuron in the excitatory population.
We construct self-consistent equations for the excitatory firing rates
at positions $\theta_{i}$ as in \eqref{mfeq-selfconsist}. However,
these now will depend additionally on the inhibitory firing rate $\nu_{I}$.
Also, the voltage-dependence of the differential equations leads to
an additional self-consistent equation for the mean-voltage $\bar{V}$.
For any position $\theta$, the excitatory self-consistent equations
are of the form (see \subsecref{Derivation-of-input-output-spike}
in Methods for detailed expressions):
\begin{align}
g(\theta) & =F(\text{input}_{\theta}\left[g\right],\nu_{I},\bar{V}(\theta))\nonumber \\
 & \equiv F\left(\int_{-\pi}^{\pi}d\varphi w(\varphi-\theta)\psi\left(g(\varphi)\right),\nu_{I},\bar{V}(\theta)\right),\label{eq:spike-selfcons-rate}\\
\bar{V}(\theta) & =G\left(\text{input}_{\theta}\left[g\right],\nu_{I},\bar{V}(\theta)\right).\label{eq:spike-selfcons-volt}
\end{align}

The function $\psi(g)$ expresses the mean synaptic activation under
presynaptic Poisson spiking at rate $g$. For accuracy, we chose to
measure $\psi$ numerically for the model of nonlinear NMDA conductance
of the recurrent excitatory synapses given in the network (see \ref{subsec:Derivation-of-input-output-spike}
in Methods).

To constrain the free parameters of $g(\theta)$, we again pick $4$
points $\theta_{i}\in\left\{ \theta_{1},\dots,\theta_{4}\right\} $,
each now yielding 2-dimensional error functions 
\begin{align}
Err_{i} & =\left(\begin{array}{c}
g(\theta_{i})-F\left(\text{input}_{\theta_{i}}\left[g\right],\nu_{I},\bar{V}(\theta_{i})\right)\\
\bar{V}(\theta_{i})-G\left(\text{input}_{\theta_{i}}\left[g\right],\nu_{I},\bar{V}(\theta_{i})\right)
\end{array}\right).\label{eq:spike-selfcons-err}
\end{align}

The resulting 8 equations are optimized for the $4$ parameters of
the parametrization $g$, as well as the additional $4$ variables
$\bar{V}(\theta_{i})$. The inhibitory population, on the other hand,
is assumed to be homogeneous. Its activity can be described by a single
mean firing rate $\nu_{I}$ and the average voltage in the inhibitory
population $\bar{V}_{I}$, resulting in a pair of additional self-consistency
errors that constrain these two variables:
\begin{align}
Err_{I}= & \left(\begin{array}{c}
\nu_{I}-F(\text{input}_{I},\nu_{I},\bar{V}_{I})\\
\bar{V}_{I}-G(\text{input}_{I},\nu_{I},\bar{V}_{I})
\end{array}\right),\label{eq:spike-selfcons-err-I}
\end{align}
where $\text{input}_{I}=\frac{1}{2\pi}\int_{-\pi}^{\pi}d\varphi\psi\left(g(\varphi)\right)$
is the mean recurrent excitatory input to inhibitory neurons.

As mentioned above, in a certain range of parameters the spiking system
possesses two dynamically stable states (\figref{4}B): the uniform
state and the ``evoked'' bump state. In this bistable regime, the
associated self-consistency Equations (\ref{eq:spike-selfcons-rate})-(\ref{eq:spike-selfcons-volt})
must have an an additional unstable solution \citep{Strogatz2000}.
Even for parameters in which the bump-state is the only stable state
of the system, the uniform state will still be a (unstable) solution
of the self-consistent equations. Accordingly, numerical solutions
of the errors Eqs. (\ref{eq:spike-selfcons-err})-(\ref{eq:spike-selfcons-err-I})
sometimes converge to the uniform state or an unstable intermediate
solution, even if a stable bump state at higher firing rates exists.
In the following we consider only the solutions with the highest spatial
modulation found under repeated solutions (see Discussion).

\subsubsection{\label{subsec:Prediction-of-firing}Prediction of firing rate profiles
from network properties}

\begin{figure}
\includegraphics[width=1\linewidth]{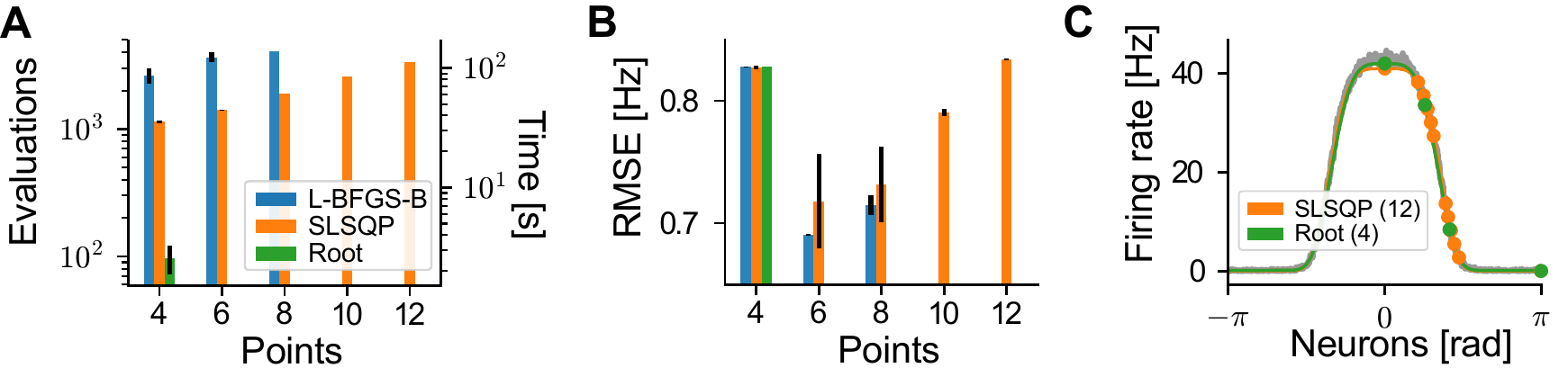}\caption[Efficient optimization of self-consistency errors]{\textbf{\label{fig:4-bottom}Efficient optimization of self-consistency
errors}. \textbf{A:} Evaluations of self-consistency errors until
convergence for three different optimization methods ($w_{+}=2.5$)
($10$ repetitions, errors show $95\%$ CI). \emph{L-BFGS-B} and \emph{SLSQP}
both minimize the sum of squared errors, \emph{Root }minimizes the
error vector directly. Right-hand axis shows wall-clock times of optimization
procedures. \textbf{B:} Averaged root mean square error (RMSE) between
optimized firing rate profiles and mean firing rates measured from
simulations ($w_{+}=2.5$, data from \figref{4}C,D,E). RMSE was calculated
for all ($800$) neurons ($n=10$, as in panel A). \textbf{C: }Optimized
firing rate profiles, together with placement of sampling points for
two combinations of optimizers and points (Root on $4$ points, SLSQP
on $12$ points). Gray line shows mean firing rate measured from simulations
($w_{+}=2.5$, as in \figref{4}C,D,E). }
\end{figure}
Above, we derived error functions constraining the parametrization
of firing rate profiles for spiking networks \eqref{spike-selfcons-err}-(\ref{eq:spike-selfcons-err-I}).
Here, we use these to predict the dependence of the spatial shape
of the firing rate profile on of a bifurcation parameter $w_{+}$,
which is the maximal strength of recurrent excitatory connections.
At $w_{+}=1$ the connection profile is homogeneous, while at larger
values local connections are stronger (\figref{4}C). The strength
of long range connections is calculated by a normalization condition
(see \subsecref{Network-connectivity} in Methods).

As $w_{+}$ is increased (\figref{4}C) above a critical value, a
spatially inhomogeneous bump state appears in simulations of the spiking
network (\figref{4}E). Our theory predicts this dependence of the
network state on the bifurcation parameter, while approximating to
a large degree the changing shape of the rate profile as the parameter
is increased (\figref{4}D,E). The firing rates of the inhibitory
population and their increase with the parameter $w_{+}$ are also
well described (\figref{4}D, black dots and lines).

As mentioned above, the error functions \eqref{spike-selfcons-err}
could, in principle, be evaluated at an arbitrary number of points.
To constrain the $4$ parameters of the parametrization $g(\theta)$,
we chose only the minimal number of points. This reduces the necessary
number of evaluations of the errors $\text{Err}_{i}$ (\figref{4-bottom}A).
Further, since in this case the dimensions of the optimization variables
$\left\{ g_{0},g_{1},g_{\sigma},g_{r}\right\} $ and the error vector
coincide, application of a more efficient numerical optimization method
(\emph{Root}, see \subsecref{Optimization-of-self-consistent}) allows
for faster optimization (\figref{4-bottom}A, green bar), which reduces
the needed time from $\sim30s$ to close to $2s$ (\figref{4-bottom}A,
right hand axis). We observe, however, that adding additional points
does slightly influence the resulting prediction (\figref{4-bottom}B),
where more sampling points placed in the flanks of the bump tend to
reduce slightly the predicted maximal firing rates (\figref{4-bottom}C,
orange points).

In a second experiment, we show that the theory can be used to efficiently
predict the effect that changing network parameters have on the shape
of the resulting firing rate profile. Similar to the simulated experiment
in \citep[Fig. 3]{Murray2012}, we systematically reduced the strength
of recurrent excitatory-to-excitatory ($g_{\text{EE}}$) and inhibitory-to-excitatory
($g_{\text{EI}}$) conductances of the network from the baseline of
the network presented in \figref{4}. Such changes of the ratio of
excitation to inhibition have been hypothesized to occur under cortical
disinhibition observed in schizophrenia \citep{Murray2012,Marin2012a}.
Recovering the result presented in the study, we see that the width
of the bump profile\footnote{Note that the network presented here generally has a wider profile
than the one investigated in \citep{Murray2012}.} depends mostly on the ratio of recurrent conductances, and thus undergoes
significant widening under disinhibition. As we have shown in \figref{4-bottom}A,
the optimization procedure for each datapoint is comparatively fast
and thus enables these type of parameter scans for wide ranges of
values under many parameters.

These results show that our approach can be used to accurately describe
the firing rate profiles of bump-attractor networks of recurrently
connected spiking excitatory and inhibitory neurons, across a range
of parameters. While evaluating the error function at more points
can lead to slightly increased accuracy of the prediction, the impact
on optimization performance is significant, increasing the needed
time by an order of magnitude.

\subsubsection{\label{subsec:Optimization-of-network}Optimization of network parameters}

\begin{figure}
\includegraphics[width=1\linewidth]{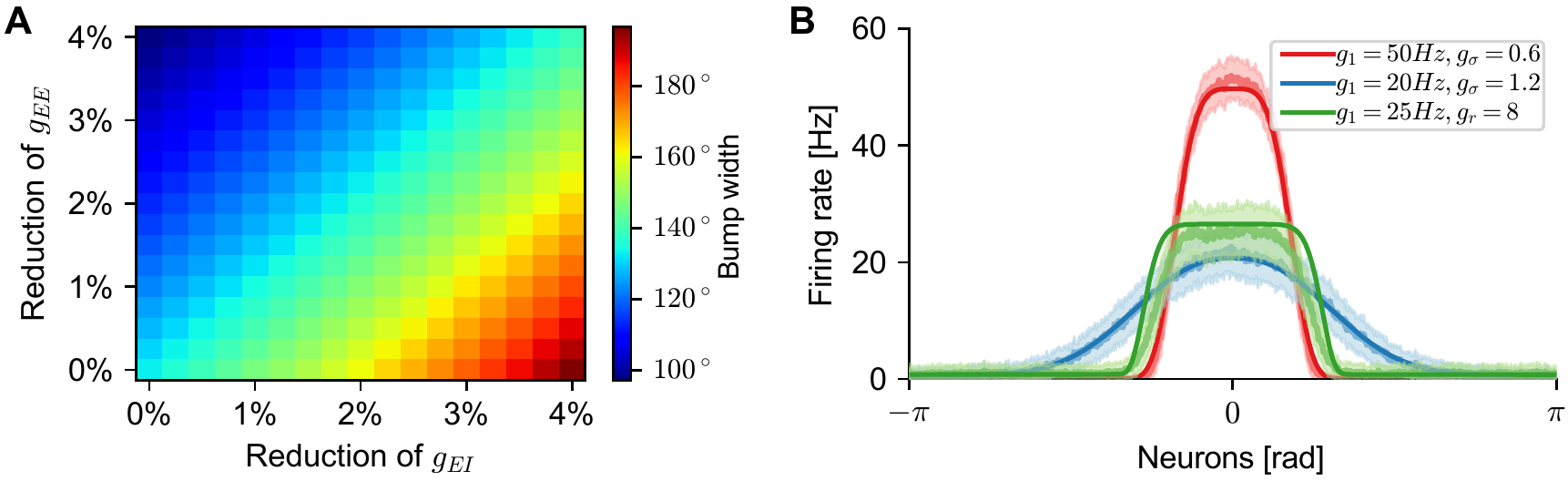}\caption[Applications: Shape prediction over parameter ranges / Optimization
of network parameters]{\textbf{\label{fig:5}Applications: Shape prediction over parameter
ranges / Optimization of network parameters}. All panels use the spiking
network model of \figref{4}. \textbf{A }Theoretically predicted width
of the firing rate profiles (full width, as predicted by the parameter
$2\cdot g_{\sigma}$) under varying strengths of recurrent excitatory-to-excitatory
($g_{\text{EE}}$) and inhibitory-to-excitatory ($g_{\text{EI}}$)
synaptic connections. Compare to the similar experiment performed
in spiking network simulations in \citet[Fig. 3]{Murray2012}. The
bifurcation parameter was kept fixed at $w_{+}=2.5$.\textbf{ B} Network
parameters optimized for 3 different shapes of the stable firing rate
profile (see \tabref{optimization-pars} for details). Solid dark
lines are theoretical predictions using the same method as in \subsecref{Prediction-of-firing}.
Lighter lines and shaded areas are mean $\pm0.5\text{STD}$ of firing
rates recorded over $3.5s$ of delay activity in 5 repetitions. }
\end{figure}
As demonstrated above for a rate-based network, our low-dimensional
approximation of continuous attractors allows the optimization of
network parameters. Here, we demonstrate that this approach extends
well to continuous attractors implemented in recurrently connected
spiking neural networks with nonlinear synaptic transmission. As in
the case of the rate network, this is achieved by fixing some desired
properties of the firing rate profile, while minimizing the error
functions Eqs. (\ref{eq:spike-selfcons-err})-(\ref{eq:spike-selfcons-err-I})
with respect to several network parameters. Here, we included the
shape parameters $w_{\sigma}$ and $w_{+}$ of the distance-dependent
connectivity, as well as the strengths of all recurrent synaptic connections:
$g_{\text{EE}},\,g_{\text{IE}},\,g_{\text{EI}},\,g_{\text{II}}$ (cf.
\tabref{optimization-pars} for details).

To find networks that admit a given shape of the firing rate profile,
we first fixed the firing rate modulation $g_{1}$ and width $g_{\sigma}$,
while optimizing the connectivity parameters $w_{+},w_{\sigma}$ as
well as the strength of all recurrent excitatory and inhibitory transmission.
In total, there were $17$ variables (see \tabref{optimization-pars}
for listings and optimization results) which were optimized (see \subsecref{Optimization-of-self-consistent}
for details). Varying these free parameters allows us to optimize
the remaining parameters of the spiking networks through a range of
shapes of the stable firing rate profile, from rather thin bumps of
high activity (\figref{5}, red) to wide bumps of low activity (\figref{5},
blue). To check whether we could optimize the spiking network to show
saturated flat-top shapes at low firing rates, we fixed $g_{1}=25Hz$
and the sharpness parameter to a high value $g_{r}=8$, while optimizing
the parameter $g_{\sigma}$ (\figref{5}, green).

Similar optimization results for all three bump shapes were achieved
by imposing additionally a low basal excitatory firing rate $\nu_{E}^{\text{basal}}$,
which constrains firing rates in the uniform (spatially homogeneous)
state for $w_{+}=1$. For the optimizations above, the basal rates
were unconstrained and varied between $1Hz$ and $5Hz$. Thereby,
the number of optimization variables is increased to $20$, since
additionally a basal inhibitory firing rate and basal inhibitory and
excitatory mean voltages need to be introduced to calculate the self-consistency
error of the basal firing rate $\nu_{E}^{\text{basal}}$. It should
be noted, that the value of this constraint affects the possible bump
shapes: for example, setting $\nu_{E}^{\text{basal}}=1Hz$ did not
yield a converging optimization for the blue and green curves of \figref{5}
\textendash{} this could be alleviated by relaxing to the higher value
$\nu_{E}^{\text{basal}}=3Hz$.

\section{Discussion}

We have presented a framework for the approximation of self-sustained
firing-rate profiles in continuous attractor neural networks. Analytical
computation of the steady states of continuous attractors is often
not possible, since it involves the solution of high-dimensional systems
of nonlinear equations, making numerical solutions necessary. Moreover,
the spatially inhomogeneous firing rate profiles of these networks
prohibit dimensional reduction of these equations by separation of
neurons into homogeneous populations, as is usually done in mean-field
approaches \citep{Brunel2001,Amit1997}. Here, we propose a simple
approach, consisting in approximating the continuous firing rate profiles
by a family of functions $g$ with only $4$ parameters. These parameters
are constrained by equations expressing the microscopic dynamics of
the neurons and synapses involved in the model, and can be optimized
to find admissible solutions. As we have shown, this can be used for
the efficient mapping of the effects that different network parameters
have on the bump shape. Next to predicting the emergent steady states
of attractors, the utility of the low-dimensional approximation is
that the derived self-consistent equations are efficiently optimizable:
we were able to use standard numerical methods to constrain the parameters
of spiking networks to show desired firing-rate profiles.

In the main text we have formulated our approach as generally as possible,
to emphasize that the method does not rely on a specific neuronal
(rate or spiking) model, as long as a prediction of firing rates given
the synaptic input can be derived. Therefore, the theory could be
extended easily to other neuron models, or connectivities where connection
probabilities are distance-dependent (e.g. \citep{Hansel2013}). The
approach presented here could also be used to predict the activity
of two-dimensional attractor models implemented in ``sheets'' of
neurons, that are often used in the context of hippocampal networks
\citep{Itskov2011,Tsodyks1999,Samsonovich1997a}. Assuming isotropy
of the connectivity (if connection strengths depend only on the Euclidean
distance between neurons), a two-dimensional generalized Gaussian
function with $g_{0},g_{1},g_{r}$ and a single parameter $g_{\sigma}$
as before could be used to approximate activity states. For non-isotropic
networks, further width parameters could be introduced and constrained
by sampling at additional points.

We have shown that our approach is amenable to the inclusion of synaptic
nonlinearities like saturating NMDA transmission (which is captured
by the synaptic activation function $\psi$, cf. \eqref{spike-selfcons-rate}).
Other sources of nonlinear synaptic transmission, for example the
activity-dependent short-term plasticity of synapses \citep{Zucker2002}
that is often investigated in the context of working memory models
\citep{Itskov2011a,Hansel2013,Mongillo2008}, can be similarly incorporated:
calculation (or numerical estimation) of a compound function $\psi$
that describes the steady-state values of synaptic activation under
all nonlinear processes affecting synaptic transmission would suffice
to adapt the theory\footnote{See Chapter 4 of \citep{Seeholzer2017}, where we apply the same method
to networks with short-term synaptic plasticity.}. Similar to the estimation of the mean-voltage in spiking networks
demonstrated here, the theory could also be extended to incorporate
adaptation effects on the steady-state firing rates of neurons \citep{Brette2005}.

For the prediction of firing rates in spiking networks we have adapted
a theory for the description of mean-field firing rates of conductance
based integrate-and-fire neurons \citep{Brunel2001} to predict spatial
firing rate profiles in recurrently connected networks \citep{Compte2000}.
Mean firing rates and mean voltages in this theory are generally expectation
values over ensembles of neurons that can be assumed to have homogeneous
activity. Strictly speaking, here we violate this assumption by taking
the rate prediction of theoretical ensembles as the prediction of
the firing rates (and mean voltages) of single neurons at given positions
along the ring attractor. However, since we are investigating the
stationary state of networks, the mean firing rates calculated from
this theory can also be interpreted as the time averaged firing rate
of single neurons \citep{Brunel1998}. Further, the approximation
of the recurrent synaptic inputs in the steady state usually relies
on the averaging over presynaptic ensembles of homogeneously firing
neurons. Nonetheless, the theory still works quite well, which might
be due to the fact that we are calculating the synaptic drive as an
integral over a continuum of presynaptic neurons, thereby effectively
averaging out deviations of the synaptic drive that are to be expected
in single neuron samples from such ensembles. 

Although bump attractor states are interesting from a functional point
of view, they are not the only solutions to ring-like networks. As
we have seen, spiking networks can show bi-stability, in which both
a stable uniform state and a stable evoked bump-state co-exist. Here,
we have neglected solutions of our theory that converged to the uniform
state (or intermediate unstable solutions) if a bump-state at larger
firing rates was also found as a solution. While it goes beyond the
scope of the current work, our theory could be extended to also predict
the dynamical stability of the states that are found: similar to approaches
in networks that admit a splitting into discrete populations \citep{Amit1997}
one could measure the magnitude of perturbations to the firing rates
at the points $\theta_{i}$ resulting from perturbations to the parametrization:
the general perturbation $g(g_{0},g_{1},g_{\sigma},g_{r})\rightarrow g(g_{0}+\delta g_{0},g_{1}+\delta g_{1},g_{\sigma}+\delta g_{\sigma},g_{r}+\delta g_{r})$
would translate into both perturbations of the rates $g(\theta_{i})=\nu_{0}(\theta_{i})+\delta\nu(\theta_{i})$
and the prediction $F(\text{input}_{\theta_{i}}\left[g_{0}\right],\bar{V}(\theta_{i}))+\delta F(\theta_{i})$.
Stability can then be determined by comparing the scale of the input
perturbations $\delta\nu(\theta_{i})$ to the predicted output perturbations
$\delta F(\theta_{i})$ \citep{Amit1997}. It is worth noting, that
calculating such linear perturbations will involve the derivative
of the synaptic activations $\psi$ (cf. \eqref{spike-selfcons-rate}),
which will also have to be estimated numerically in cases where no
analytical formula for $\psi$ can be found.

Our choice of parametrization of the firing rate profile was heuristic,
guided by the shapes observed in numerical simulations. The framework
presented here, though, could be used with any other family of functions
parametrized by a small number of parameters, which should be adapted
to the shapes to be approximated. For example, multi-modal ring-attractor
profiles resulting from narrower connectivities (see e.g. \citep{Laing2002},
or \citep{Wei2012} for a spiking network similar to the one investigated
here) can not be approximated by the unimodal family chosen here.
Since the topology of ring-attractors is periodic, a natural candidate
for such a generalization would be the family of finite Fourier series.
However, the nonzero frequency components necessary to faithfully
approximate shapes that deviate far from simple (e.g. cosine-shaped)
unimodal distributions might require a large number of Fourier coefficients
for parametrization. 

In this report we mostly chose as many positions $\theta_{i}$ along
the attractor manifold as there are free parameters of the profile
$g$ to be constrained. In principle, the number of free parameters
can be chosen independently of the number of positions, by performing
numerical optimization on a dimension-agnostic sum of squared errors.
We have shown that matching the error dimension to the number of parameters
permits using efficient optimization methods that significantly speed
up the optimization. However, we have also investigated under-determined
(see \figref{5}B) and over-determined (see \figref{4-bottom}A-C)
systems, which also converged to similar solutions. Finally, when
optimizing the network parameters for desired spatial profiles (\figref{3}
and \figref{5}), choosing optimization goals outside the space of
possible solutions of the network dynamics did not allow the procedure
to converge. Thus, our approach could be used to estimate the boundaries
of the solution space for a given neural network, by starting the
optimization at a known solution and varying the shape parameters
until convergence fails.

\section{Methods}

\subsection{Rate network model\label{subsec:Rate-network}}

We study a network of $N=100$ recurrently connected rate neurons
indexed by $i\in\{0,...,N-1\}$, where each neuron $i$ is described
by a single variable $\nu_{i}(t)$ which denotes the firing rate of
the neuron \citep{Miller2012}. 

The neuronal firing rate is given by a nonlinear input-output function
$F$ of a synaptic variable $s_{i}$ (assuming that the membrane time
constant is considerably faster than the synaptic variable):
\begin{align}
\nu_{i}(t) & =F(s_{i}(t))=\frac{\nu_{\text{max}}}{2}\left[1+\tanh\left(\frac{s_{i}(t)}{s_{0}}\right)\right].\label{eq:rate-neurons}
\end{align}
 Here, $\nu_{\text{max}}=50\text{Hz}$ sets the maximal firing frequency,
and $s_{0}=1$ is the (dimensionless) scale of the synaptic input.

The input to neuron $i$ is mediated through the synaptic variable
$s_{i}$:
\begin{align}
\dot{s}_{i}(t) & =-\frac{s_{i}(t)}{\tau_{s}}+\sum_{j=0}^{N-1}w_{ij}\nu_{j}(t),\label{eq:rate-input}
\end{align}
where $\dot{s}(t)=\frac{d}{dt}s(t)$ denotes the temporal derivative,
$\tau_{s}=100ms$ is the synaptic time constant, and $w_{ij}$ are
the recurrent connection weights (see below), and $0\leq i\leq N-1$.

Neurons are organized at circular positions $\theta_{i}=i\cdot\frac{2\pi}{N}-\pi\in[-\pi,\pi)$
with identified boundaries, such that neuron $0$ is the direct neighbor
of neuron $N-1$. The recurrent connections depend only on the distance
between neurons in the resulting angular space: the connection $w_{ij}$
from neuron $j$ to neuron $i$ is given by a generalized Gaussian
function, with 4 free parameters controlling its shape:
\begin{align}
w_{ij} & =w(\theta_{i}-\theta_{j})=\frac{1}{N}\left(w_{0}+w_{1}\exp\left[-\left(\frac{\left|\min\left(\left|\theta_{i}-\theta_{j}\right|,2\pi-\left|\theta_{i}-\theta_{j}\right|\right)\right|}{w_{\sigma}}\right)^{w_{r}}\right]\right).\label{eq:rate-connectivity}
\end{align}

The parameters $\{w_{0},w_{1},w_{\sigma},w_{r}\}$ used for the networks
of \figref{2} and \figref{3} are given in \tabref{weights-systems}.

\subsection{\label{subsec:Spiking-network-methods}Spiking network model}

Spiking simulations are based on a reimplementation of a popular ring-attractor
model of visuospatial working memory \citep{Compte2000} in the NEST
simulator \citep{Diesmann2007}. Parameters were modified from the
original publication to produce the results shown in \figref{4},
\figref{4-bottom}, and \figref{5} (see \tabref{parameters} for
parameter values). For completeness we restate the definition of the
model here.

\subsubsection{Neuron model\label{subsec:Neuron-model}}

Neurons are modeled by leaky integrate-and-fire dynamics with conductance
based synaptic transmission \citep{Compte2000,Brunel2001}. The network
consists of recurrently connected populations of $N_{E}$ excitatory
and $N_{I}$ inhibitory neurons, both additionally receiving external
spiking input with spike times generated by $N_{\text{ext}}$ independent,
homogeneous Poisson processes, with mean rates $\nu_{ext}$. Following
\citep{Compte2000}, we assume that external excitatory inputs are
mediated by fast AMPA receptors, while recurrent excitatory currents
are mediated only by slower NMDA channels.

The neuronal dynamics for neurons in both excitatory and inhibitory
populations are governed by the following system of differential equations
indexed by $i\in\{0,...,N_{E/I}-1\}$ (with different sets of parameters
for each population):
\begin{eqnarray}
C_{m}\dot{V}_{i}(t) & = & -I_{i}^{L}(t)-I_{i}^{\text{Ext}}(t)-I_{i}^{\text{I}}(t)-I_{i}^{\text{E}}(t),\label{eq:base_deqs}\\
I_{i}^{P} & = & g_{P}\,s_{i}^{P}(V_{i}(t),t)\,\left(V_{i}(t)-V_{P}\right),\nonumber 
\end{eqnarray}
where $P\in\{\text{\text{L,Ext,I,E}}\}$. Here, $C_{\text{m}}$ is
the membrane capacitance and $V_{\text{L}},V_{\text{E}},V_{\text{I}}$
are the reversal potentials for leak, excitatory currents, and inhibitory
currents, respectively. The parameters $g_{P}$ for $P\in\{\text{\text{L,Ext,I,E}}\}$
are fixed scales for leak (L), external input (Ext) and recurrent
excitatory (E) and inhibitory (I) synaptic conductances, which are
dynamically gated by the (possibly voltage dependent) gating variables
$s_{i}^{P}(V,t)$. In the main text we refer to the conductance scales
of excitatory neurons by the ``strength of synaptic connections''
$g_{\text{EE}}=g_{\text{E}}$ and $g_{\text{EI}}=g_{\text{I}}$. Similarly,
for inhibitory neurons we refer to the conductance scales by the ``strengths''
$g_{\text{IE}}=g_{\text{E}}$ and $g_{\text{II}}=g_{\text{I}}$. The
gating variables $s_{i}^{P}$ are described in detail below, however
we set the leak conductance gating variable to $s_{i}^{L}=1$. 

The model neuron dynamics (\eqref{base_deqs}) are integrated until
their voltage reaches a threshold $V_{\text{thr}}$. At any such time,
the respective neuron emits a spike and its membrane potential is
reset to the value $V_{\text{res}}$. After each spike, voltages are
clamped to $V_{\text{res}}$ for a refractory period of $\tau_{\text{ref}}$
(see \tabref{parameters} for parameter values).

\subsubsection{Synaptic gating variables\label{subsec:Synaptic-gating-variables}}

The synaptic gating variables $s_{i}^{P}(t)$ for $P\in\{\text{\text{Ext,I}}\}$
for external and inhibitory currents are exponential traces of the
firing times $t_{j}$ of all presynaptic neurons $j$:
\begin{equation}
\dot{s}_{i}^{P}(t)=-\frac{s_{i}^{P}(t)}{\tau_{P}}+\sum_{j\in\text{pre}(P)}\sum_{t_{j}}\delta\left(t-t_{j}\right),\label{eq:syn_linear_filter}
\end{equation}
where the sum runs over all neurons presynaptic to the neuron $i$
regarding the connection $P$.

For the recurrent excitatory gating variables $s_{i}^{\text{E}}$
a nonlinear NMDA model is used \citep{Wang2001}. This model has second
order kinetics for NMDA channel activation \citep{Destexhe1994},
which result in a saturation of channels. Together with a voltage
dependence $\text{Mg}(V_{i})$ of the conductance (due to the release
of the $\text{Mg}^{2+}$ block, see \citep{Jahr1990}) this yields
the following dynamics:
\begin{eqnarray}
s_{i}^{\text{E}}(V,t) & = & \text{Mg}(V_{i})\sum_{j=1}^{N_{E}}w_{ij}^{\text{E}}y_{j}(t),\label{eq:deq-nmda-full}\\
\dot{y}_{j} & = & -\frac{y_{j}}{\tau_{\text{E}}}+\alpha x_{j}(t)\left(1-y_{j}\right),\label{eq:deq-nmda2}\\
\dot{x}_{j} & = & -\frac{x_{j}}{\tau_{\text{E,rise}}}+\sum_{t_{j}}\delta\left(t-t_{j}\right),\label{eq:deq-nmda3}\\
\text{Mg}\left(V\right) & = & \frac{1}{1+\gamma\exp\left(-\beta V\right)}.\label{eq:Mg-NMDA}
\end{eqnarray}

See \tabref{parameters} for parameter values used in simulations.

\subsubsection{Network connectivity\label{subsec:Network-connectivity}}

All connections except for the recurrent excitatory connections are
all-to-all and uniform. The recurrent excitatory connections are chosen
to be distance-dependent. As in the rate model, each neuron of the
excitatory population with index $i\in\{0,...,N_{E}-1\}$ is assigned
an angular position $\theta_{i}=i\cdot\frac{2\pi}{N_{E}}-\pi\in[-\pi,\pi)$.
Recurrent excitatory NMDA connections $w_{ij}^{E}$ from neuron $j$
to neuron $i$ are then given by the Gaussian function $w^{E}(\theta)$:
\begin{align*}
w_{ij}^{E} & =w^{E}(\theta_{i}-\theta_{j})=w_{0}+\left(w_{+}-w_{0}\right)\exp\left(-\left[\min\left(\left|\theta_{i}-\theta_{j}\right|,2\pi-\left|\theta_{i}-\theta_{j}\right|\right)\right]^{2}\frac{1}{2\sigma_{w}^{2}}\right).
\end{align*}

Additionally, for each neuron we keep the integral over all recurrent
connection weights normalized, resulting in the normalization condition
$\frac{1}{2\pi}\int_{-\pi}^{\pi}d\varphi w^{E}(\varphi)=1.$ This
normalization ensures that varying the maximum weight $w_{+}$ will
not change the total recurrent excitatory input if all excitatory
neurons fire at the same rate. Here, we choose $w_{+}$ as a free
parameter and constrain the baseline connection weight to
\begin{align*}
w_{0} & =\frac{w_{+}\sigma_{w}\text{erf}\left(\frac{\pi}{\sqrt{2}\sigma_{w}}\right)-\sqrt{2\pi}}{\sigma_{w}\text{erf}\left(\frac{\pi}{\sqrt{2}\sigma_{w}}\right)-\sqrt{2\pi}}.
\end{align*}

\subsection{\label{subsec:Mean-field-equations}Self-consistent equations}

\subsubsection{\label{subsec:Placement-of-sampling}Placement of sampling points}

Self consistent equations (\eqref{mfeq-selfconsist}) are constructed
for both rate-based and spiking neurons by using the low-dimensional
parametrization \eqref{g-gen-gaussian} described in the main text.
As mentioned there, we choose the top of the firing rate profile $\theta_{1}=0$,
as well as the lowest point $\theta_{4}=\pi$. For the intermediate
points $0<\theta_{i}<\pi$ for $i\in\{2,3\}$, we sample the firing
rate profile by inverting the function $g$ to give a sample at a
desired height $h_{i}=g_{0}+a_{i}\left(g_{1}-g_{0}\right)$ with $0<a_{i}<1$.
This yields a relation for the position which depends on the shape
parameters $g_{r}$ and $g_{\sigma}$:
\[
\theta_{i}=-g_{\sigma}\log\left(a_{i}\right)^{\frac{1}{g_{r}}}.
\]

For all figures except for \figref{4-bottom} and \figref{5}, the
intermediate points were chosen by setting $a_{2}=0.2$ and $a_{3}=0.8$,
although we show (\figref{2} and \figref{3}) that the exact choice
of points affects the solutions only slightly. 

In \figref{4-bottom} we iterate through even numbers $p\geq4$ of
sampling points. As before, we choose $\theta_{1}=0$, as well as
the lowest point $\theta_{p}=\pi$. Generalizing the placement of
$4$ points described above, the remaining $p-2$ points were chosen
as $\theta_{k}=\nicefrac{0.4}{k}$ for $1<k\leq\nicefrac{p}{2}$,
and $\theta_{k}=1-\nicefrac{0.4}{k}$ for $\nicefrac{p}{2}<k\leq p$.
For the optimization of spiking network parameters shown in \figref{5},
we chose $7$ sampling points: we first chose $p=6$ points by the
scheme just described, then added the point $\theta_{7}=0.5$.

\subsubsection{\label{subsec:Derivation-of-input-output-rate}Derivation of input-output
functions for the rate network}

For the rate network, we set \eqref{rate-input} to zero and solve
for the steady-state input $s_{i}$, which yields
\begin{align*}
s_{i} & =\tau_{s}\sum_{j=0}^{N-1}w_{ij}\nu_{j}=\tau_{s}\sum_{j=0}^{N-1}w_{ij}g(\theta_{j})\\
 & \approx\frac{\tau_{s}N}{2\pi}\int_{-\pi}^{\pi}d\varphi w(\theta_{i}-\varphi)g(\varphi).
\end{align*}

Here, we have replaced the activity of neurons in the network by our
parametrization $g$. In the second line we approximated the summation
$\frac{1}{N}\sum_{j=0}^{N-1}$ by the integral $\frac{1}{2\pi}\int_{-\pi}^{\pi}d\varphi$
and exploited that the connectivity $w$ is only dependent on the
angular distance between the neuron $i$ (at position $\theta_{i}$)
and neurons $j$ (at varying positions $\theta_{j}=\varphi$), to
replace $w_{ij}\rightarrow w(\theta_{i}-\varphi)$. We use this steady
state input in \eqref{rate-neurons} to arrive at \eqref{rate-error}.

\subsubsection{\label{subsec:Derivation-of-input-output-spike}Derivation of input-output
functions for the spiking network}

In the rate model presented in \subsecref{Rate-network} the firing
rates are given by \eqref{rate-neurons}. In the spiking network,
we have to approximate the expected firing rates of neurons. To this
end, we first replace the synaptic activation variables $s^{P}(V,t)$
for $P\in\{I,E,\text{ext}\}$ by their expectation values under Poisson
input. For the linear synapses this yields 
\begin{align*}
\left\langle s^{\text{ext}}(t)\right\rangle _{\text{Poisson}}= & \tau_{\text{ext}}\nu_{\text{pre}},\\
\left\langle s^{\text{I}}(t)\right\rangle _{\text{Poisson}}= & \tau_{\text{I}}\nu_{I}.
\end{align*}

The nonlinear synaptic activation of NMDA synapses under stimulation
with Poisson processes at rates $\nu_{j}$ was estimated by simulating
Eqs. (\ref{eq:deq-nmda2})-(\ref{eq:deq-nmda3}) under varying presynaptic
firing rates and fitting an interpolating function to the temporal
means of the synaptic activation $\psi(\nu_{j})\equiv\langle y_{j}\rangle_{t}$.
An analytical approximation of the function $\psi(\nu_{j})$ was stated
in \citep[p. 80]{Brunel2001}. We instead chose to numerically fit
this function to simulated data, since for higher firing rates the
analytical approximation tended to over-estimate the synaptic activations. 

We then define the expected recurrent excitatory input, assuming presynaptic
Poisson firing, by 
\begin{align}
J_{i} & \equiv\frac{1}{N_{E}}\sum_{i=0}^{N_{E}-1}w_{ij}^{E}\psi(\nu_{j}).\label{eq:total-input}
\end{align}

Following \citep{Brunel2001}, we linearize the voltage dependence
\eqref{Mg-NMDA} at the mean voltage $\left\langle V\right\rangle $
and reduce the differential equations of \eqref{base_deqs} to dimensionless
form. The resulting expressions depend only on the mean firing rates
and mean voltages of excitatory and inhibitory neurons (see \subsecref{Detailed-derivation-of}
of the Appendix for the detailed expressions and derivations):
\begin{eqnarray}
\tau_{i}\dot{V}_{i} & = & -\left(V_{i}-V_{L}\right)+\mu_{i}+\sigma_{i}\sqrt{\tau_{i}}\eta_{i}(t)\label{eq:reduced-voltage-deq}\\
\mu_{i} & = & \mu_{i}\left(J_{i},\nu_{I},\nu_{\text{ext}},\left\langle V_{i}\right\rangle \right)\nonumber \\
\sigma_{i} & = & \frac{g_{\text{ext}}}{C_{m}}\left(\left\langle V\right\rangle -V_{E}\right)\tau_{\text{ext}}\sqrt{\tau_{i}N_{\text{ext}}\nu_{\text{ext}}}.\nonumber \\
\tau_{i} & = & \tau_{i}\left(J_{i},\nu_{I},\nu_{\text{ext}},\left\langle V_{i}\right\rangle \right)\nonumber \\
\left\langle \eta_{i}(t)\right\rangle  & = & 0\nonumber \\
\left\langle \eta_{i}(t)\eta_{i}(t')\right\rangle  & = & \frac{1}{\tau_{\text{ext}}}\exp(-\frac{\left|t-t'\right|}{\tau_{\text{ext}}})\nonumber 
\end{eqnarray}

Here, $\mu_{i}$ is the bias of the membrane potential due to synaptic
inputs, and $\sigma_{i}$ measures the scale of fluctuations in the
membrane potential due to random spike arrival approximated by the
Gaussian process $\eta_{i}$. Due to active synaptic conductances,
the effective membrane time constant $\tau_{i}$ is decreased from
the intrinsic membrane time-constant $\nicefrac{C_{m}}{g_{L}}$ \textendash{}
its value thus depends on all presynaptic firing rates (and the mean
voltage, see \subsecref{Detailed-derivation-of} of the Appendix).

The prediction $F$ of the mean firing rates and $\left\langle V_{i}\right\rangle $
of mean voltages of populations of neurons governed by this type of
differential equation can be well approximated by \citep{Brunel2001}
(see also the published corrections in \citep{Brunel2001a}):
\begin{eqnarray}
\phi\left[\mu_{i},\sigma_{i},\tau_{i}\right] & = & \left(\tau_{\text{ref}}+\sqrt{\pi}\tau_{i}\int_{\beta(\mu_{i},\sigma_{i})}^{\alpha(\mu_{i},\sigma_{i})}du\exp(u^{2})\left[1+\text{erf}\left(u\right)\right]\right)^{-1},\label{eq:siegert}\\
\alpha(\mu_{i},\sigma_{i}) & = & \frac{V_{\text{reset}}-V_{L}-\mu_{i}}{\sigma_{i}}\left(1+\frac{\tau_{\text{ext}}}{2\tau_{i}}\right)+1.03\sqrt{\frac{\tau_{\text{ext}}}{\tau_{i}}}-\frac{\tau_{\text{ext}}}{\tau_{i}},\label{eq:alpha}\\
\beta(\mu_{i},\sigma_{i}) & = & \frac{V_{\text{reset}}-V_{L}-\mu_{i}}{\sigma_{i}},\label{eq:beta}\\
\left\langle V_{i}\right\rangle  & = & \mu_{i}+V_{L}-\left(V_{\text{thr}}-V_{\text{reset}}\right)\phi\left[\mu_{i},\sigma_{i},\tau_{i}\right]\tau_{i}.\label{eq:siegert_voltage}
\end{eqnarray}

As in the rate model, we first replace the network activity $\nu_{j}$
on the right hand side of \eqref{total-input} by our parametrization
$g(\theta_{j})$. We then approximate the summation $\frac{1}{N_{E}}\sum_{j=0}^{N_{E}-1}$
with an integral  $\frac{1}{2\pi}\int_{-\pi}^{\pi}d\varphi$, and
replace the connectivity by its continuous equivalent $w_{ij}\rightarrow w(\theta_{i}-\varphi)$
to arrive at:
\begin{align*}
J_{i} & \approx\frac{1}{2\pi}\int_{-\pi}^{\pi}d\varphi w^{E}\left(\theta_{i}-\varphi\right)\psi(g\left(\varphi\right))\\
 & \equiv\frac{1}{2\pi}\text{input}_{\theta_{i}}\left[g\right].
\end{align*}

We then substitute this relation in Eqs. (\ref{eq:siegert}) and (\ref{eq:siegert_voltage})
to arrive at 
\begin{align}
\mu_{i} & =\mu_{i}\left(\text{input}_{\theta_{i}}\left[g\right],\nu_{I},\nu_{\text{ext}},\left\langle V_{i}\right\rangle \right),\nonumber \\
\tau_{i} & =\tau_{i}\left(\text{input}_{\theta_{i}}\left[g\right],\nu_{I},\nu_{\text{ext}},\left\langle V_{i}\right\rangle \right),\nonumber \\
F(\text{input}_{\theta_{i}}\left[g\right],\nu_{I},\left\langle V_{i}\right\rangle ) & \equiv\phi\left[\mu_{i},\sigma_{i},\tau_{i}\right],\nonumber \\
G(\text{input}_{\theta_{i}}\left[g\right],\nu_{I},\left\langle V_{i}\right\rangle ) & \equiv\mu_{i}+V_{L}-\left(V_{\text{thr}}-V_{\text{reset}}\right)g_{i}(\theta_{i})\tau_{i},\label{eq:self-consistent-eqs}
\end{align}
which defines Eqs. (\ref{eq:spike-selfcons-rate}) and (\ref{eq:spike-selfcons-volt})
of the main text. 

\subsubsection{\label{subsec:Optimization-of-self-consistent}Optimization of self-consistent
equations}

For each point $\theta_{i}$ that we choose to sample from the excitatory
population, the theory of \subsecref{Derivation-of-input-output-spike}
yields $2$ constraining Equations  (\ref{eq:siegert}) and (\ref{eq:siegert_voltage}).
The inhibitory population, being homogeneous and unstructured, yields
$2$ equations, for the 2 free variables $\nu_{I}$ and $\left\langle V_{i}\right\rangle $.
Since we choose a low-dimensional parametrization for the excitatory
population, the number of free variables increases only by $1$ (the
mean voltage $\left\langle V_{i}\right\rangle $) for each point $\theta_{i}$
that we choose to evaluate, while yielding the same $2$ constraining
equations. This allows us to choose at minimum $4$ evaluation points
to constrain the $4$ free parameters of the parametrization (see
\tabref{optimization-pars} for a listing).

The errors $\text{Err}_{i}$ (and $\text{Err}_{I}$, for spiking networks)
between firing rate predictions and the firing rate parametrization
are numerically minimized using methods provided in the \emph{Scipy}
package \citep{Oliphant2007}. In particular, if the dimension of
the error function matches the number of parameters, we are able to
use the efficient \emph{optimize.root} solver (\emph{Root} in the
main text), which applies a modified version of the Powell hybrid
method \citep{Powell1970}, but does not provide constraints on valid
parameter regions. Here, we implemented artificial constraints by
returning a high error for dimensions that leave the bounded region.
The same optimization results were achieved by using the slower \emph{optimize.minimize}
method, which allows optimization (of the sum of squared errors $\text{SSE}=\sum_{i}\text{Err}_{i}^{2}$,
or $\text{SSE}=\sum_{i}\text{Err}_{i}^{2}+\text{Err}_{I}^{2}$ for
spiking networks) in constrained parameter regions via the L-BFGS-B
\citep{Byrd1995} and\emph{ }SLSQP \citep{Kraft1988}. For spiking
networks, we normalized firing rate errors by the firing rate $\nu_{\text{max}}=100Hz$
and voltage differences by the voltage range $V_{\text{thres}}-V_{\text{reset}}$,
to ensure comparable contributions to the SSE for variables with different
dimensions. 

For the optimization results of \figref{5} we chose $7$ sampling
points (see \subsecref{Placement-of-sampling} for details), which
yielded $16$ errors, including those of the inhibitory population.
These were used to optimize $17$ free parameters using the SLSQP
algorithm (see \tabref{optimization-pars} for a listing). We also
tried using $8$ sampling points, which brings both the number of
equations and free parameters up to $20$ \textendash{} this yielded
similar results at increased processing time (the possibly faster
\emph{Root} solver failed to converge most of the time).

Wall clock times for error functions in \figref{4-bottom}A were measured
on a single core of a MacBook Pro with 2,6 GHz Intel Core i5 processor,
using the Python benchmark \emph{timeit.timeit} (minimum wall clock
time of $100$ repetitions). We first measured average time for evaluation
of a single error $\text{Err}_{i}$ of \eqref{spike-selfcons-err},
which evaluated to $t_{E}=4.59ms$ (100 repetitions of 10 executions).
For a single evaluation of the inhibitory error $\text{Err}_{I}$
of \eqref{spike-selfcons-err-I} we found $t_{I}=0.98ms$ (the numerical
integration performed in the calculation of $\text{input}_{\text{I}}=\frac{1}{2\pi}\int d\varphi\psi\left(g\left(\varphi\right)\right)$
is faster). The wall clock time $T$ (see \figref{4-bottom}A, right
axis) for a given number $n$ of error vector evaluations on $p$
points was then calculated by $T=n\left(p\cdot t_{E}+t_{I}\right)$.

\subsection{Spiking simulations}

All network simulations where performed in the NEST simulator \citep{Diesmann2007}
using fourth-order Runge-Kutta integration as implemented in the GSL
package \citep{Galassi2009}. For the simulation results shown in
\figref{4} and \figref{5}B, networks underwent a transient initial
period of $t_{\text{initial}}$. Neurons centered at a position of
$\theta=0$ then received a short and strong excitatory input mediated
by additional Poisson firing onto AMPA receptors ($500ms,2\text{kHz}$)
with connections scaled down by a factor of $g_{\text{signal}}=0.5$.
The external input ceased at $t=t_{\text{off}}$. 

Simulations were run until $t=t_{\text{max }}$ and spikes were recorded
and converted to firing rates by spike counts in a $75ms$ window
shifted at a time resolution of $1ms$. For every time step, the firing
rates across the whole population were then rectified (by measuring
the phase of the first spatial Fourier coefficient and setting it
to $\theta=0$ by rotation of the angular space) to center the bump
of activity around the position $\varphi=0$. The resulting centered
firing rates were then sampled at an interval of $60ms$ in the interval
$\left[t_{\text{off}}+500ms,t_{\text{max}}\right]$ for 5 repetitions
of the network simulation with the same microscopic parameters. In
\figref{4}B times were: $t_{\text{initial}}=1.8s,t_{\text{off }}=2s,t_{\text{max}}=5s$.
In \figref{4}D,E we chose: $t_{\text{initial}}=0.5s,t_{\text{off }}=1s,t_{\text{max}}=3s$.
For \figref{5}B: $t_{\text{initial}}=0.5s,t_{\text{off }}=1s,t_{\text{max}}=5s$.

\section{Acknowledgements}

\bibliographystyle{unsrturl}
\phantomsection\addcontentsline{toc}{section}{\refname}\bibliography{bibliography}
\newpage{}

\appendix

\part*{Appendix}

\section{\label{subsec:Detailed-derivation-of}Detailed derivation of dimensionless
voltage equations}

In this section we give details on the derivation of \eqref{reduced-voltage-deq}
as well as the resulting full expressions. This closely follows \citep[pp. 79--81]{Brunel2001},
while keeping a slightly simplified notation. 

We first replace all synaptic activations in \eqref{base_deqs} by
their expected values under Poisson input, which also introduces the
expected recurrent excitatory input $J_{i}$ (cf. \eqref{total-input}):
\begin{align*}
s_{i}^{\text{ext}}(t) & \rightarrow N_{\text{ext}}\tau_{\text{ext}}\nu_{\text{pre}}+\Delta_{S,ext},\\
s_{i}^{\text{I}}(t) & \rightarrow N_{I}\tau_{\text{I}}\nu_{I},\\
s_{i}^{E}(t) & \rightarrow\text{Mg}(V_{i})\sum_{i=0}^{N_{E}-1}w_{ij}^{E}\psi(\nu_{j})=\text{Mg}(V_{i})N_{E}J_{i}.
\end{align*}

Here, $\Delta_{S,ext}$ represents fluctuations of around the mean
of $s_{i}^{\text{ext}}$ due to random spike arrival at fast AMPA
synapses. Since the synaptic timescales (GABA, NMDA) of the other
synaptic activations are much longer, these fluctuations can be neglected.
We then rearrange \eqref{base_deqs} to dimensionless form, which
yields:
\begin{eqnarray*}
\frac{C_{m}}{g_{L}}\dot{V}_{i} & = & -\left(V_{i}-V_{L}\right)\left[1+T_{I}\nu_{I}+T_{\text{ext}}\nu_{\text{ext}}+\frac{g_{E}}{g_{L}}\text{Mg}(V_{i})N_{E}J_{i}\right]\\
 &  & +\left(V_{I}-V_{L}\right)T_{I}\nu_{I}+\left(V_{E}-V_{L}\right)\left[T_{\text{ext}}\nu_{\text{ext}}+\frac{g_{E}}{g_{L}}\text{Mg}(V_{i})N_{E}J_{i}\right]\\
 &  & +\frac{g_{\text{ext}}}{g_{L}}\left(V_{i}-V_{E}\right)\Delta_{S,\text{ext}},
\end{eqnarray*}
where $T_{\text{ext}}=N_{\text{ext}}\tau_{\text{ext}}\frac{g_{\text{ext}}}{g_{L}},\,T_{I}=N_{I}\tau_{I}\frac{g_{I}}{g_{L}}$
are effective timescales of external and inhibitory input.

To get rid of the nonlinear voltage dependence of the right hand side
through $\text{Mg}(V_{i})$, we linearize this function (cf. \eqref{Mg-NMDA})
around the mean voltage $\left\langle V_{i}\right\rangle $:
\begin{eqnarray*}
\frac{V_{i}-V_{E}}{1+\gamma\exp(-\beta V_{i})} & = & \frac{\left\langle V_{i}\right\rangle -V_{E}}{\rho}+\left(V-\left\langle V\right\rangle \right)\frac{\rho+\beta\left(\left\langle V_{i}\right\rangle -V_{E}\right)\left(\rho-1\right)}{\rho^{2}},
\end{eqnarray*}
where $\rho=1+\gamma\exp\left(-\beta\left\langle V_{i}\right\rangle \right)$. 

After replacing the voltage dependence in the fluctuation term by
the mean voltage, we arrive at
\begin{eqnarray}
\frac{C_{m}}{g_{L}}\dot{V}_{i} & = & -\left(V_{i}-V_{L}\right)\left[1+T_{I}\nu_{I}+T_{\text{ext}}\nu_{\text{ext}}+\left(\rho_{1}+\rho_{2}\right)J_{i}\right]\nonumber \\
 &  & +\left(V_{I}-V_{L}\right)T_{I}\nu_{I}+\left(V_{E}-V_{L}\right)\left[T_{\text{ext}}\nu_{\text{ext}}+\rho_{1}J_{i}\right]\nonumber \\
 &  & +\rho_{2}\left(\left\langle V_{i}\right\rangle -V_{L}\right)J_{i}+\frac{g_{\text{ext}}}{g_{L}}\left(\left\langle V_{i}\right\rangle -V_{E}\right)\Delta_{S,\text{ext}}.\nonumber \\
\rho_{1} & = & \frac{g_{\text{E}}N_{E}}{g_{L}\rho}\nonumber \\
\rho_{2} & = & \beta\frac{g_{\text{E}}N_{E}\left(\left\langle V_{i}\right\rangle -V_{E}\right)\left(\rho-1\right)}{g_{L}\rho^{2}}\nonumber \\
\rho & = & 1+\gamma\exp\left(-\beta\left\langle V_{i}\right\rangle \right).\label{eq:dimless_form}
\end{eqnarray}
Finally, we replace the fluctuations $\Delta_{S,ext}$ by independent
Gaussian noise processes with zero mean $\left\langle \eta_{i}(t)\right\rangle =0$
and simpler autocorrelation $\left\langle \eta_{i}(t)\eta_{j}(t')\right\rangle =\frac{1}{\tau_{\text{ext}}}\exp(-\frac{\left|t-t'\right|}{\tau_{\text{ext}}})\delta_{ij}$,
to arrive at the full form of \eqref{reduced-voltage-deq} in the
main text:
\begin{eqnarray}
\tau_{i}\dot{V}_{i} & = & -\left(V_{i}-V_{L}\right)+\mu_{i}+\sigma_{i}\sqrt{\tau_{i}}\eta_{i}(t)\label{eq:app-full-voltage-eq}\\
S_{i} & = & 1+T_{I}\nu_{I}+T_{\text{ext}}\nu_{\text{ext}}+\left(\rho_{1}+\rho_{2}\right)J_{i}\nonumber \\
\mu_{i}S_{i} & = & \left(V_{I}-V_{L}\right)T_{I}\nu_{I}+\left(V_{E}-V_{L}\right)T_{\text{ext}}\nu_{\text{ext}}+\nonumber \\
 &  & \left[\rho_{1}\left(V_{E}-V_{L}\right)+\rho_{2}\left(\left\langle V\right\rangle -V_{L}\right)\right]J_{i}\nonumber \\
\sigma_{i} & = & \frac{g_{\text{ext}}}{C_{m}}\left(\left\langle V\right\rangle -V_{E}\right)\tau_{\text{ext}}\sqrt{\tau_{i}N_{\text{ext}}\nu_{\text{ext}}}.\nonumber \\
\tau_{i} & = & \frac{C_{m}}{g_{L}S_{i}}\nonumber 
\end{eqnarray}

Reducing the conductance based differential equation \eqref{base_deqs}
of the main text to the simplified form \eqref{app-full-voltage-eq},
now allows us to compute the mean firing rate as a functions of the
(input-like) bias $\mu_{i}$ and fluctuation term $\sigma_{i}$, according
to \eqref{siegert} of the last section.

\begin{table}
\resizebox{1.0\textwidth}{!}{%
\begin{tabular}{lcccc}
\toprule 
Parameter name & Parameter Symbol  & Units & Excitatory neurons  & Inhibitory neurons \tabularnewline
\midrule 
Neuron number & $N_{\bullet}$ & 1 & $N_{E}=800$  & $N_{I}=200$ \tabularnewline
Poisson neuron number  & $N_{\text{ext}}$  & 1 & $1000$  & $1000$ \tabularnewline
Membrane capacitance & $C_{m}$  & pF & $500$  & $200$ \tabularnewline
Exc. reversal potential & $V_{E}$  & mV & $0$  & $0$ \tabularnewline
Inh. reversal potential & $V_{I}$  & mV & $-70$  & $-70$ \tabularnewline
Leak reversal potential & $V_{L}$  & mV & $-70$  & $-70$ \tabularnewline
After spike reset potential & $V_{\text{res}}$  & mV & $-60$  & $-60$ \tabularnewline
Spiking threshold & $V_{\text{thr}}$  & mV & $-50$  & $-50$ \tabularnewline
NMDA parameter 1 & $\beta$  & 1 & $0.062$  & $0.062$ \tabularnewline
NMDA parameter 2 & $\gamma$  & 1 & $1/3.57$  & $1/3.57$ \tabularnewline
NMDA rise parameter & $\alpha$ & s & 0.5 & 0.5\tabularnewline
External conductance & $g_{ext}$  & nS & $2.08$  & $1.62$ \tabularnewline
Recurr. inh. conductance & $g_{\text{I}}$  & nS & $1.336$  & $1.024$ \tabularnewline
Recurr. exc. conductance & $g_{\text{E}}$  & nS & $0.381$  & $0.292$ \tabularnewline
Leak conductance & $g_{\text{L}}$  & nS & $25$  & $20$ \tabularnewline
External synaptic timescale  & $\tau_{\text{ext}}$  & ms & $2$  & $2$ \tabularnewline
Recurr. inh. timescale & $\tau_{I}$  & ms & $10$  & $10$ \tabularnewline
Recurr. exc. timescale & $\tau_{E}$  & ms & $100$  & $100$ \tabularnewline
Recurr. exc. rise timescale & $\tau_{\text{E,rise}}$  & ms & $2$  & $2$ \tabularnewline
Membrane time constant $\frac{C_{m}}{g_{L}}$ & $\tau_{m}$  & ms & $20$  & $10$ \tabularnewline
Refractory period & $\tau_{\text{ref}}$  & ms & $2$  & $1$ \tabularnewline
Width of distance dep. weights & $\sigma_{w}$ & rad & $\frac{18\deg}{360\deg}\cdot2\pi\approx0.31$ & -\tabularnewline
Frequency of Poisson neurons & $\nu_{\text{ext}}$  & Hz & $2.4Hz$  & $2.4Hz$ \tabularnewline
\bottomrule
\end{tabular}}\caption[Parameters for spiking simulations]{\label{tab:parameters}\textbf{Parameters for spiking simulations}.
Parameter values are modified from \citep{Compte2000} and \citep{Brunel2001}.}
\end{table}
\begin{table}
\begin{tabular}{lcccc}
\toprule 
System  & $w_{0}$  & $w_{1}$ & $w_{\sigma}$ & $w_{r}$\tabularnewline
\midrule 
Sys. 0  & $-0.8$ & $2.3$ & $0.9$ & $2.0$\tabularnewline
\midrule 
Sys. 1 & $-1.0$ & $10.$ & $0.2$ & $2.0$\tabularnewline
\midrule 
Sys. 2 & $-3.0$ & $15.0$ & $0.5$ & $2.0$\tabularnewline
\bottomrule
\end{tabular}\caption[Connectivity parameters of rate models]{\label{tab:weights-systems}\textbf{Connectivity parameters of rate
models.}}
\end{table}
\begin{table}
\resizebox{1.0\textwidth}{!}{%
\begin{tabular}{llccc}
\toprule 
Figure  & Fixed & Optimized & \# Optimized & Points / Errors\tabularnewline
\midrule 
\figref{4} & $\begin{array}[t]{c}
w_{+},w_{\sigma}\\
g_{\text{EE}},g_{\text{IE}},g_{\text{EI}},g_{\text{II}}
\end{array}$ & $\begin{array}[t]{c}
g_{0},g_{1},g_{\sigma},g_{r}\\
\left\langle V\right\rangle _{\theta_{1}},\dots,\left\langle V\right\rangle _{\theta_{4}}\\
\nu_{I},\left\langle V\right\rangle _{I}
\end{array}$ & 10 & 4 / 10\tabularnewline
\midrule 
\figref{5} red & $\begin{array}[t]{c}
g_{1}=50Hz\\
g_{\sigma}=0.6\\
\\
\\
\end{array}$ & $\begin{array}[t]{c}
g_{0},g_{r}\\
w_{+}=4.100,w_{\sigma}=0.1899\\
g_{\text{EE}}=0.3923,g_{\text{IE}}=0.3958\\
g_{\text{EI}}=1.1611,g_{\text{II}}=0.9570\\
\left\langle V\right\rangle _{\theta_{1}},\dots,\left\langle V\right\rangle _{\theta_{7}}\\
\nu_{I},\left\langle V\right\rangle _{I}
\end{array}$ & 17 & 7 / 16\tabularnewline
\midrule 
\figref{5} blue & $\begin{array}[t]{c}
g_{1}=20Hz\\
g_{\sigma}=1.2\\
\\
\\
\end{array}$ & $\begin{array}[t]{c}
g_{0},g_{r}\\
w_{+}=2.423,w_{\sigma}=0.4750\\
g_{\text{EE}}=0.1798,g_{\text{IE}}=0.1858\\
g_{\text{EI}}=0.7882,g_{\text{II}}=0.7632\\
\left\langle V\right\rangle _{\theta_{1}},\dots,\left\langle V\right\rangle _{\theta_{7}}\\
\nu_{I},\left\langle V\right\rangle _{I}
\end{array}$ & 17 & 7 / 16\tabularnewline
\midrule 
\figref{5} green & $\begin{array}[t]{c}
g_{1}=25Hz\\
g_{r}=8\\
\\
\\
\end{array}$ & $\begin{array}[t]{c}
g_{0},g_{\sigma}\\
w_{+}=4.4917,w_{\sigma}=0.0909\\
g_{\text{EE}}=0.4397,g_{\text{IE}}=0.4742\\
g_{\text{EI}}=1.1933,g_{\text{II}}=1.1948\\
\left\langle V\right\rangle _{\theta_{1}},\dots,\left\langle V\right\rangle _{\theta_{7}}\\
\nu_{I},\left\langle V\right\rangle _{I}
\end{array}$ & 17 & 7 / 16\tabularnewline
\bottomrule
\end{tabular}}\caption[Parameters optimized in spiking networks]{\label{tab:optimization-pars}\textbf{Parameters optimized in spiking
networks.} For constraints in \figref{4} see the parameter values
in \tabref{parameters}. For network parameters in \figref{5}, we
additionally give the values obtained by optimization. \emph{Points}
is the number of sampling points. \emph{Errors} is the number errors
used for optimization, this includes 2 errors for the inhibitory population,
in addition to 2 errors per sampling point.}
\end{table}

\end{document}